\documentclass[onecolumn]{aa} 

\usepackage{graphicx}
\usepackage{txfonts}
\usepackage{natbib}
\begin{document}
   \title{Post-flare evolution of AR 10923 with Hinode/XRT}

   \subtitle{}

   \author{S. Parenti
          \inst{1}\fnmsep\thanks{Now at the Institut d'Astrophysique Spatiale, Univ. Paris Sud, 91405 Orsay Cedex, France}
          \and 
           F. Reale
          \inst{2,3}
           \and 
           K. K. Reeves
             \inst{4}  }

   \offprints{s.parenti@oma.be}

   \institute{Royal Observatory of Belgium, 3 Av. Circulaire, 1180 Bruxelles, Belgium \\
         \email{s.parenti@oma.be}
         \and
             Dip. di Science Fisiche ed Astronomiche, Univ. di Palermo 1, 90134 Palermo, Italy
          \and
				INAF-Osservatorio Astronomico di Palermo “G. S. Vaiana,” Piazza del Parlamento 1, 90134 Palermo, Italy.
           \and
            Harvard-Smithsonian Center for Astrophysics, 60 Garden Street MS 58, Cambridge, MA 02138}

   \date{Received ; accepted }

 
  \abstract
   {Flares are dynamic events which involve rapid changes in coronal magnetic topology end energy release. Even if they may be localized phenomena, the magnetic disturbance at their origin may propagate and be effective in a larger part of the active region.}
   {We investigate the temporal evolution of a flaring active region with respect to the loops morphology, the temperature, and emission measure distributions.}
   {We consider $Hinode/XRT$ data of a  the 2006 November 12th C1.1 flare.  We inspect the evolution of the morphology of the flaring region also with the aid of TRACE data. XRT filter ratios are used to derive temperature and emission measure maps and evolution. }
   {The analyzed flare includes several brightenings.  We identify a coherent sequence of tangled and relaxed loop structures before, during, and after the brightenings. Although the thermal information is incomplete because of pixel saturation at the flare peak, thermal maps show fine, evolving spatial structuring. Temperature and emission measure variations show up in great detail, and we are able to detect a secondary heating of larger loops close to the proper flaring region.  Finally we estimate the amount of energy released in these flaring loops during the flare decay.}


   \maketitle
%

\section{Introduction}

Solar and stellar flares are identified as impulsive events of energy release with associated sudden brightening in the full waveband spectrum. Magnetic complexity seems to be a key point for the occurring of flares, and magnetic reconnection is often invoked as the triggering process. Models of colliding and reconnecting magnetic flux tubes have been developed both for two ribbon flares and compact flares \citep[e.g.][]{sturrock84, jakimiec98}. In these models the increase of twisting of the magnetic flux tubes in the corona accumulates free magnetic energy, which is released when the flux tubes are brought into contact. The magnetic complexity may be due to flux tubes footpoint motion linked to stochastic photospheric motion \citep{farnik96}, or flux emergence \citep[e.g.][]{zirin83}. 

Soft X-rays (SXR) are the most suitable band to map the full flare-cycle from the rising to the cooling phases. This emission depends on the magnetic geometry and volume involved in the flare and on the  properties of the energy deposited. Soft X-rays are the source of information to understand the flare phenomena and its environment.
 
Diagnostics of solar and stellar flares in soft X-ray have been developed by several authors \cite[e.g][]{kopp84,pallavicini90,sylwester93,reale02, reale07b}. However, 
these methods need to be coupled to reliable techniques which invert the X-ray emission in terms of, for example, plasma temperature and emission measure (or density).

The SXR imagers delivered data only from few spectral bands which until recently  limited the diagnostic possibilities with respect to spectroscopy data  and 
made the results of the inversion process dependent on strong assumptions.
Nevertheless, the XRT instrument on board the $Hinode$ satellite (launched 2006) has brought new possibilities in terms of SXR plasma diagnostics with broad band data. The instrument contains nine different filters sensitive to coronal emission, which is the largest number ever available. With these data, \cite{reale07} developed a new technique  for temperature measurement (called combined improved filter ratio, CIFR) consisting of the ratio of multiple filters, while   \cite{reale09} showed the multi-thermal information that can be derived by combining the information of different single filter ratios. 
The instrument XRT also has  an unprecedented spatial resolution and sensitivity for this class of instruments. 

In this paper we combine the high cadence and spatial resolution multi-band information of XRT to study the temporal evolution of the topological complexity and thermal structuring in an active region prior and after a flare (Sect. \ref{sec_top}), addressing  the question of its role during flaring activity. The XRT data, here combined   with those from TRACE, provide for the first time enough spatial resolution ($1-2\arcsec$) to allow an analysis over the fine coronal spatial scales.

Here we show the first application of the new CIFR method to flare data. This, joined to the standard filter ratio, is used to study the temporal evolution of the  temperature and emission measure of loops of different sizes around the flaring region (Sect. \ref{sec_temp} and Sect.\ref{sec_em}). Conclusions and implications on the energy budget will be discussed in Sect. \ref{sec_disc}.

\section{Observations and methodology}

The X-ray Telescope (XRT) on $Hinode$ \citep{golub07} is based on grazing incidence optics using the Wolter-Schwarzschild design, which focuses the image on a CCD camera with an on-axis point spread function (PSF) of 1.0" +/- 0.12". Nine filters are located on two wheels located near the focal plane and in front of the CCD camera. These allow the user to vary the X-ray passband to obtain different  temperature responses in the range $6.1<\log~T<7.5$.

\begin{figure}[ht]
\centering
\hspace{-.5cm}\includegraphics[width=70mm]{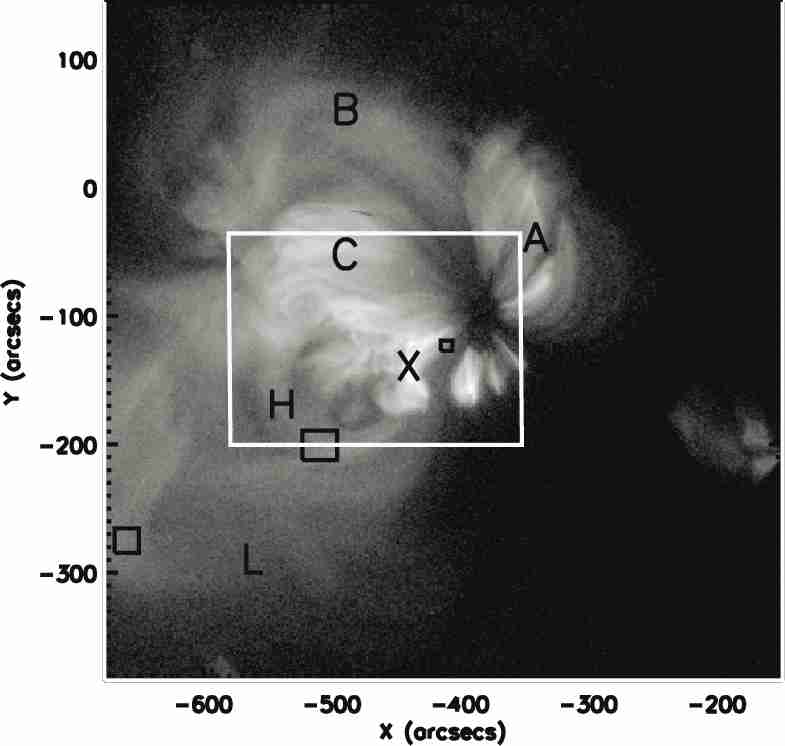}\\
\hspace{-.4cm}\includegraphics[width=70mm]{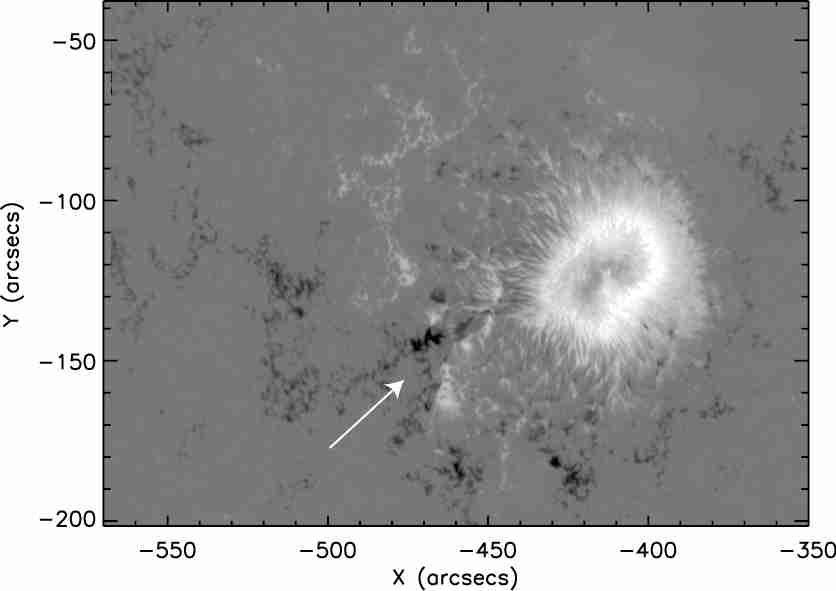}\\
\caption{$XRT/Al_{poly}$ map of active region 10923 on the  2006  November the 12 at 10:36UT (top) and the SOT magnetogram (Bottom). Labeled in black are the loop systems analyzed in this work and the boxes for the thermal analysis. The white box on the top image indicates the SOT FOV. The arrow indicates the flaring region. 
}\label{al_p}
\end{figure}

The instrument XRT followed AR 10923 during the full passage of the disk in November 2006. In the $512\arcsec \times 512\arcsec$ images taken by this instrument (Fig. \ref{al_p}, top),
the active region appears to be made of  toroidal loops of similar length around the spot and, on the south west side, by larger loops (here named L), which may have been part of the footpoints outside the XRT field of view (FOV). This region was analyzed in quiet condition on the November 12 by \cite{reale07,reale09}. 
The south west part of the AR was the location of several flares during this passage. 
Figure \ref{al_p} bottom shows the Hinode/SOT magnetogram taken at 10:40UT, which includes over a   smaller FOV compared to the upper panel the spot and its surroundings.  This figure allows us to better see the high concentration of mixed polarity in the southwest part of the image, which is also indicated by the arrow. This corresponds to the region labeled X on the XRT image (top). The neutral line pointed at by the arrow is the site of a filament, as seen in absorption in the  TRACE 171 data.

\begin{figure}[h]
\centering
\includegraphics[scale=3]{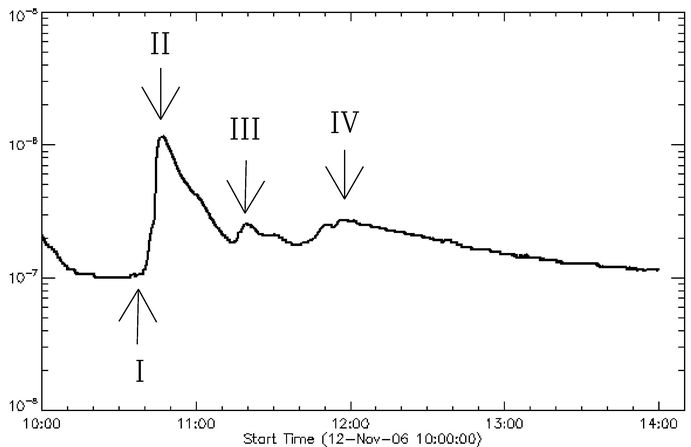}
\caption{GOES12 flux (1-8 \AA) for the flare analyzed in this work. The four main phases of the analysis are marked by the arrows: the begin of XRT data (I), the main flare peak (II), the second brightening (III), and the end of XRT data (IV).}\label{goes}
\end{figure}

 Figure \ref{goes} shows the GOES flux in the 1-8 \AA~channel where we marked 
 the main phases of the data analysis of this work by arrows.  
We analyze a C1.1 flare which started on 2006 November 12 at 10:38UT (I) and peaked at 10:46UT (II).
The peak of the flare was followed by weaker brightening, the most intense of which  saturated  the hottest XRT channels at 11:26UT (III). We stopped our analysis before 12UT (IV). 
A close investigation of the active region showed that the main changes in temperature and morphology are localized around the flaring region. We concentrate our analysis on the latter, which includes the regions $\mathrm{X}$, $\mathrm{L}$, and $\mathrm{H}$ (Fig. \ref{al_p}).

The evolution of the level of loops tangling  in the active region during the flare will be compared  to the quiet condition. To do this we also 
 selected XRT data taken on 2006 November 10 at 17:55 UT in addition to an early quiet time close to the flare at 10:36 UT. We chose this day because the active region was particularly quiet between about 4:00 and 23:00 UT. 

\begin{table*}[ht]
\caption{XRT and TRACE exposure times}
\label{table1}      
\centering          
\begin{tabular}{c c c c l l l }     
\hline\hline  
&  \multicolumn{5}{c}{XRT} &  TRACE \\                        
Time (UT)& $Al\_poly$ &$C\_poly$ & $Be\_thin$ & $Be\_med$ & $Al\_med$ & 171 \\ 
\hline                    
17:55 & 0.18 & 0.4 & 1.4 & 5.8 &11.57 &  \\
10:36 & 0.26 & 0.36& 1.44& 8.19 & 16.39 &\\
10:46 &  0.18 & 0.26 & 1.03 & 5.80 & 11.57 & \\
11:01 & 0.01 & 0.02 & 0.07 & 0.36 & 0.76 & \\ 
11:09 &      &     &     &     &      & 55.1\\
11:16 &  0.03 &0.05 & 0.18 & 1.03 & 2.05 & 32.8 \\
11:21 &  0.1 &  0.13 &  0.51 &  2.90 & 5.80 &  $\vdots$ \\
11:26 & 0.36 & 0.51 & 0.36 & 2.05 & 4.10 & $\vdots$\\
11:31  &0.26 & 0.36& 1.44& 8.19 & 16.39 & $\vdots$\\
11:46 & 0.18 & 0.51 & 1.03 & 5.80 & 11.57 & $\vdots$\\
11:56 & 0.18 & 0.18 & 0.73 & 4.10 & 8.19  & $\vdots$\\
13:46 & 0.3 & 0.4 & 1.4 & 8.2 & 16.4 & $\vdots$\\
\hline                  
\end{tabular}
\tablefoot{Start time of the observation (first column), the exposure times in seconds (second-sixth columns) for XRT and TRACE (last column).}    
\end{table*}

From 11:09 UT on the active region was also  observed by TRACE 171 with a $1024\arcsec\times 1024\arcsec$ field of view. Unfortunately, no observations are available at the time of the  flare's peak. We selected a subset of images to be compared to the XRT data.

Table \ref{table1} lists the observations used for the present analysis. The instrument XRT observed using sequences of five filters ($Al\_poly$, $C\_poly$, $Be\_thin$, $Be\_med$ (here called F4), $Al\_med$ (here called F5)) with about 5 min cadence. The table shows the variation of the exposure times along the sequence.
The FOV was of $512\arcsec \times 512\arcsec$. Figure \ref{Al_poly_time} shows the time evolution of the  area of interest through a selection of 
$Al\_poly$ filter images.

The data were processed with the $xrt\_prep$, $trace\_prep$ and $fg\_prep$ softwares available on the $SolarSoft$ package. For the temperature analysis we used the latest XRT filter thickness calibration available on the XRT package, but with a correction on the $Be\_{med}$ filter, as discussed in  \cite{reale09}.

Our temperature analysis  is based on the filter ratio method. 
Under the hypothesis of isothermal plasma along the line of sight, the ratio of filters with a different sensitivity to the temperature is a function of the latter only. If the plasma is not isothermal, as  generally expected, the temperature so derived is a quantity-weighted by the emission measure of the emitting plasma and the filter response function. Despite the strong assumption, the combined use of several different filter ratios allows the recovery of some information about the thermal structure along the line of sight. 
In the present work we use the filter ratio of F4/F5 to get the information of the hottest components detected by the instrument. 
We also use the
 the combined improved filter ratio (CIFR) method developed by \cite{reale07} to build temperature maps of the bulk temperature of the active region. We divide the geometric mean of the five filters by  that of the two softest ones. We set a threshold 
in these data which changes between 200 and 300 $\mathrm{DN/s}$  for F4, and 150 and 200 $\mathrm{DN/s}$ for F5, depending on the exposure time.

The CIFR is an improvement of a two filter method when multi-filter ($>2$) data are available. It improves the sensibility in temperatures and enhances the details of the temperature structures  at the expense of losing possible multitemperature information along the line of sight. It is also more robust to calibration errors.
For the choice of the filters, this ratio is more sensitive to the warm temperatures than the F4/F5 ratio.

\section{Data analysis}

\subsection{Morphology}
\label{sec_top}

\begin{figure*}[ht]
   \centering
\includegraphics[scale=.3]{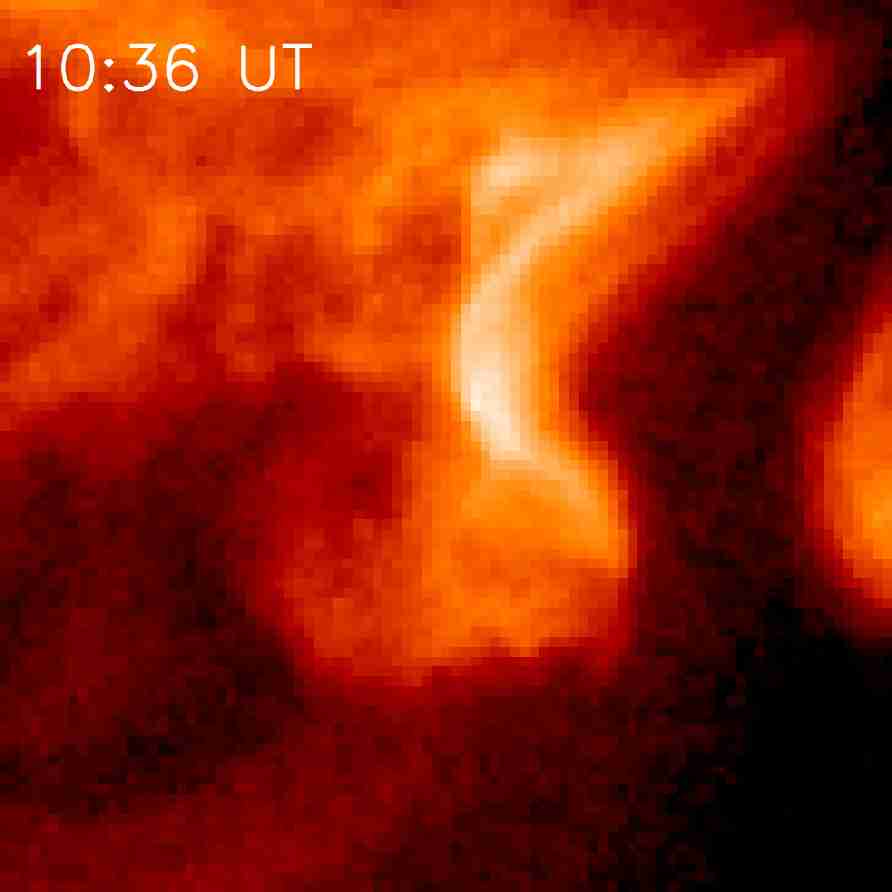}
 \includegraphics[scale=.3]{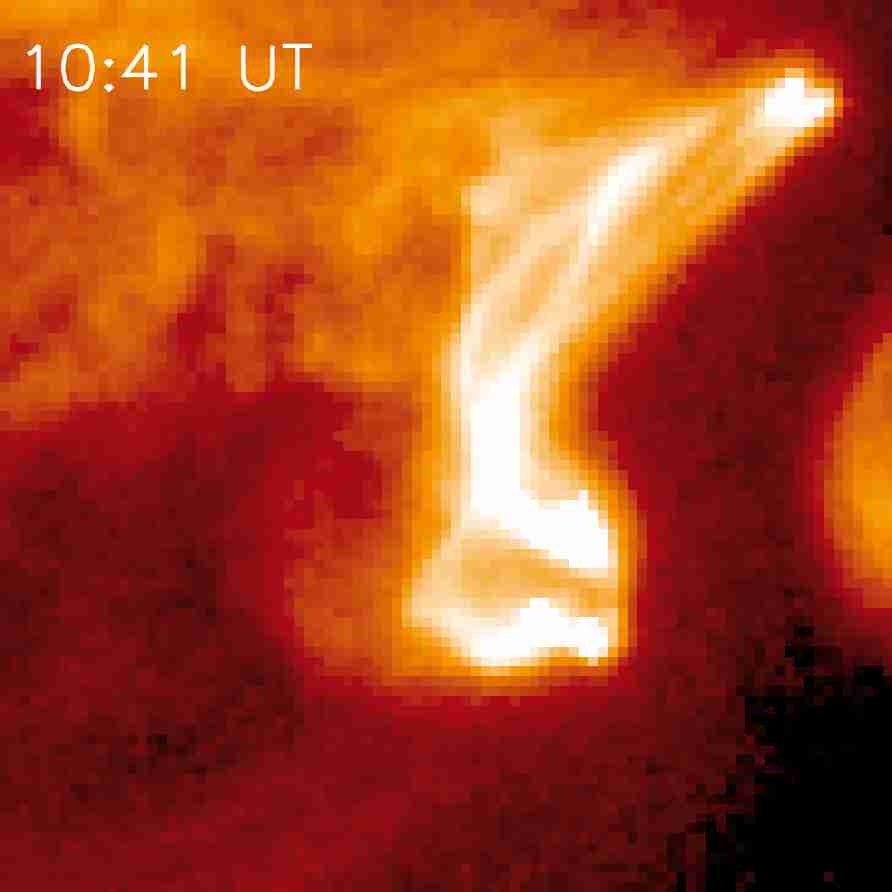}
\includegraphics[scale=.3]{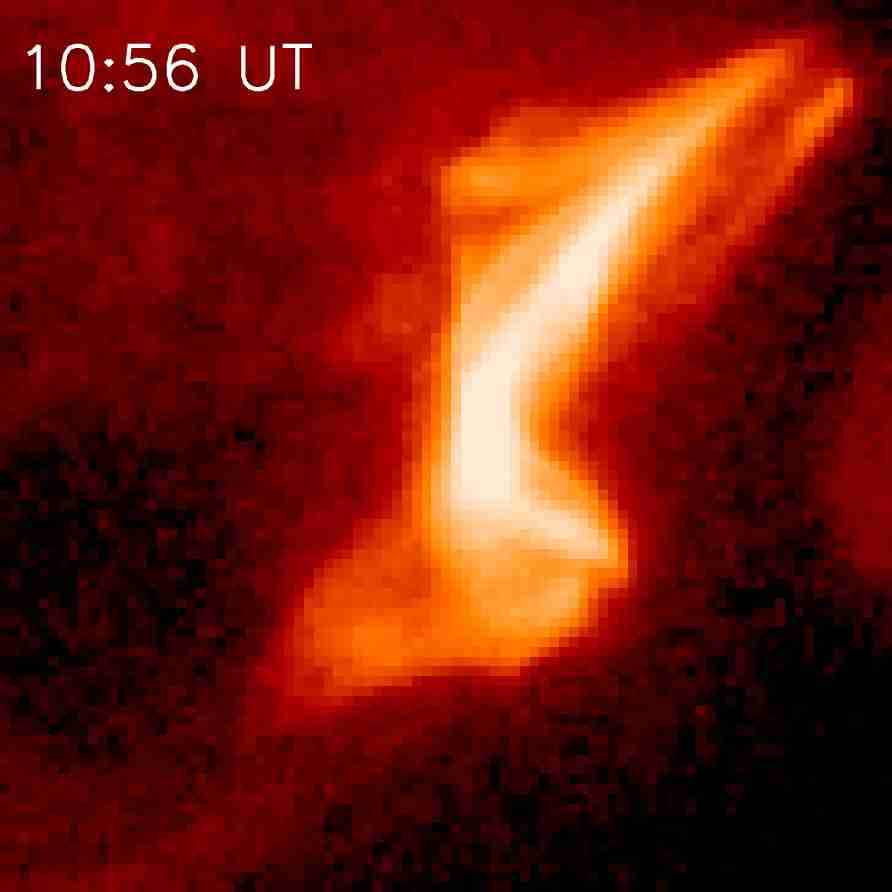}\\
 \includegraphics[scale=.3]{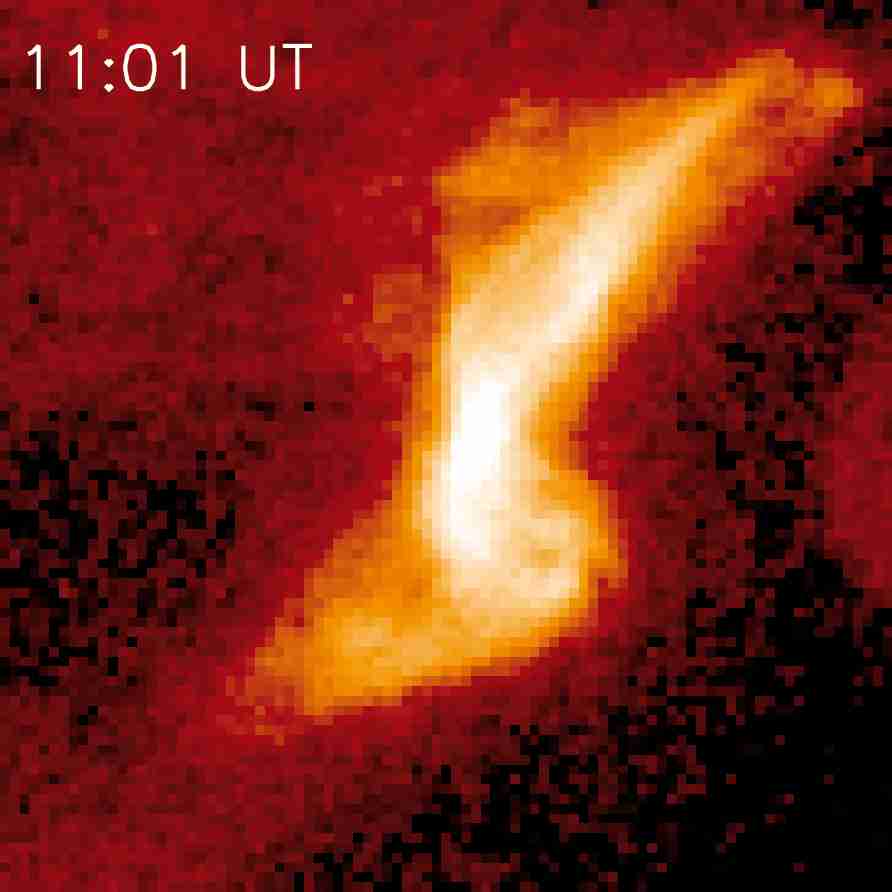}
\includegraphics[scale=.3]{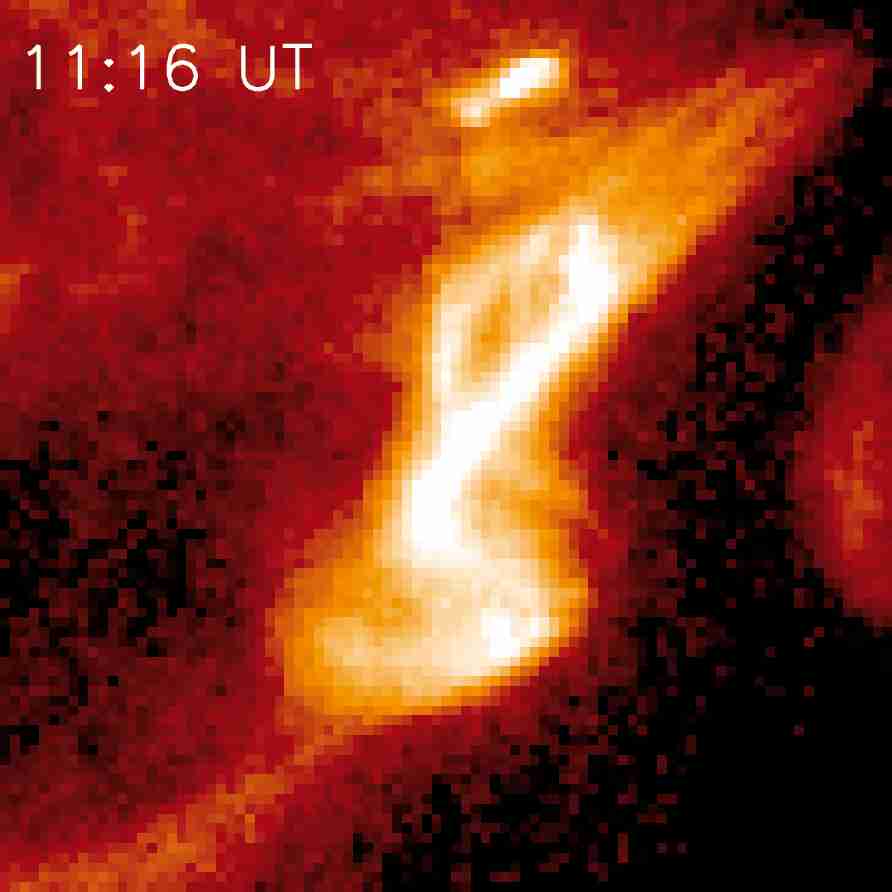}
\includegraphics[scale=.3]{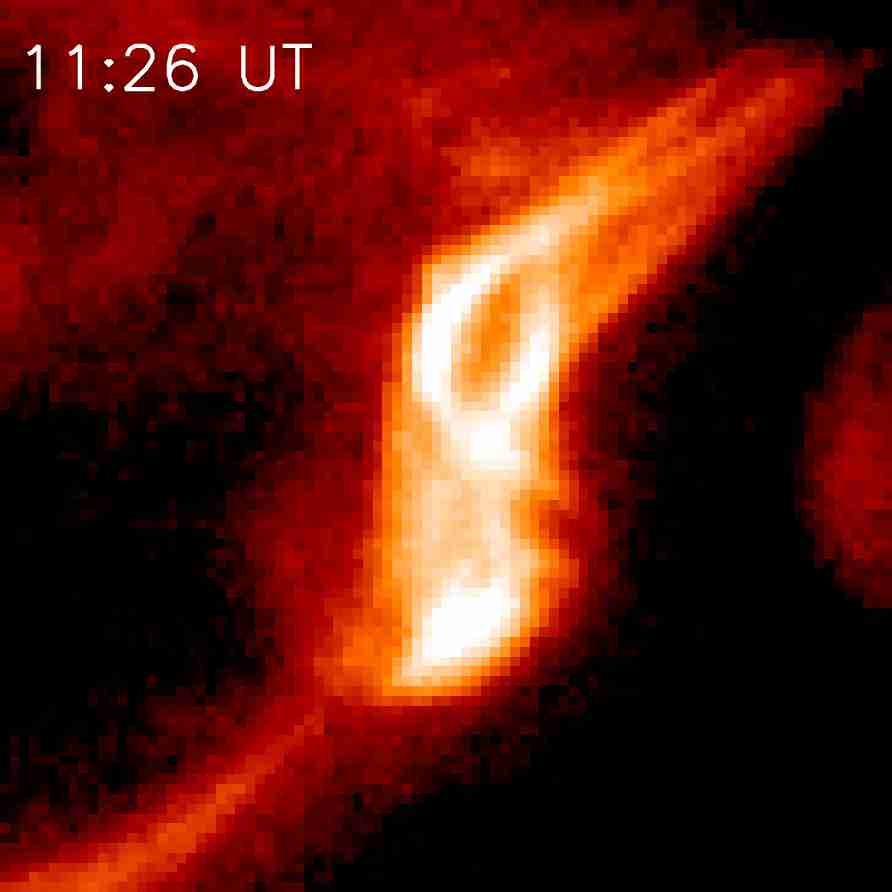}\\
\includegraphics[scale=.43]{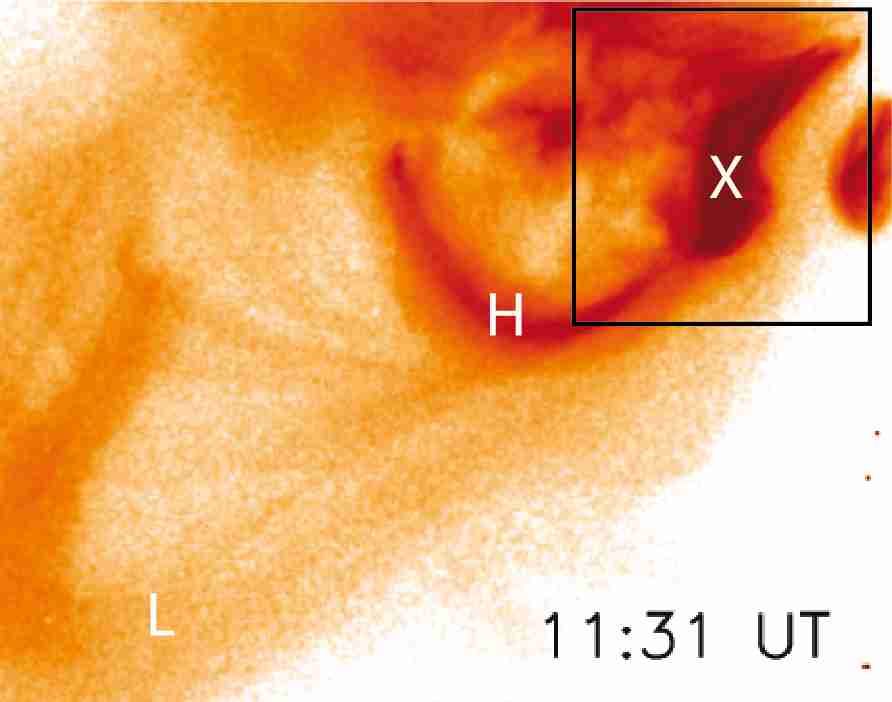}
\includegraphics[scale=.43]{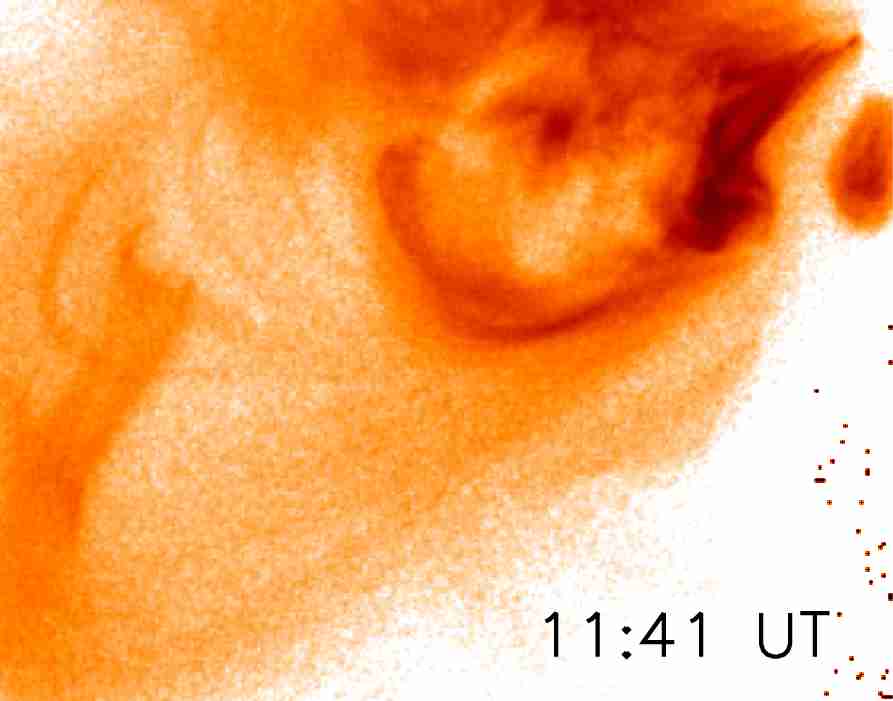}\\
\vspace{-1cm}\includegraphics[scale=.43]{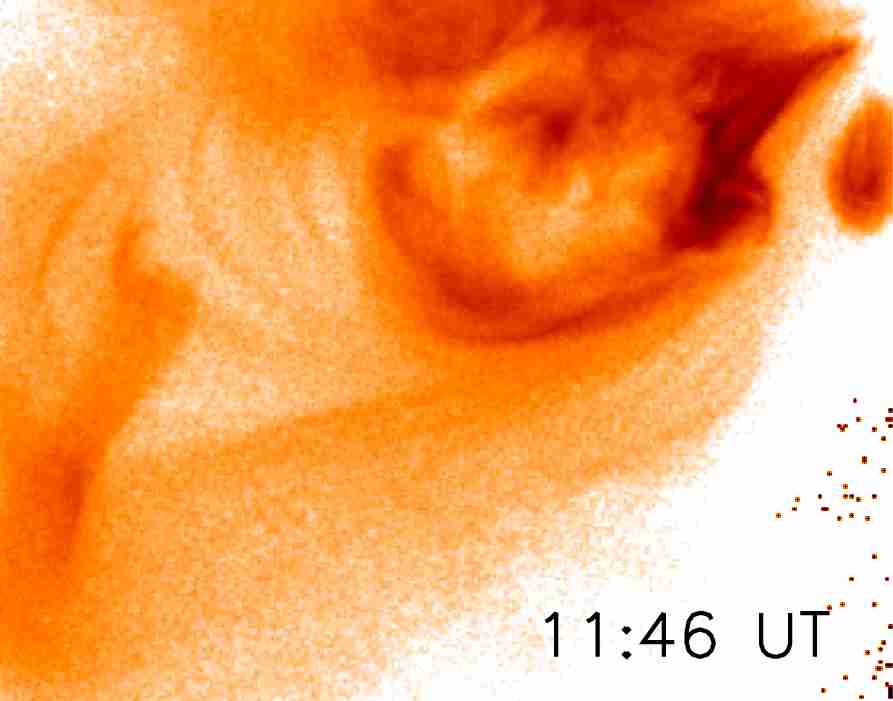}
\includegraphics[scale=.43]{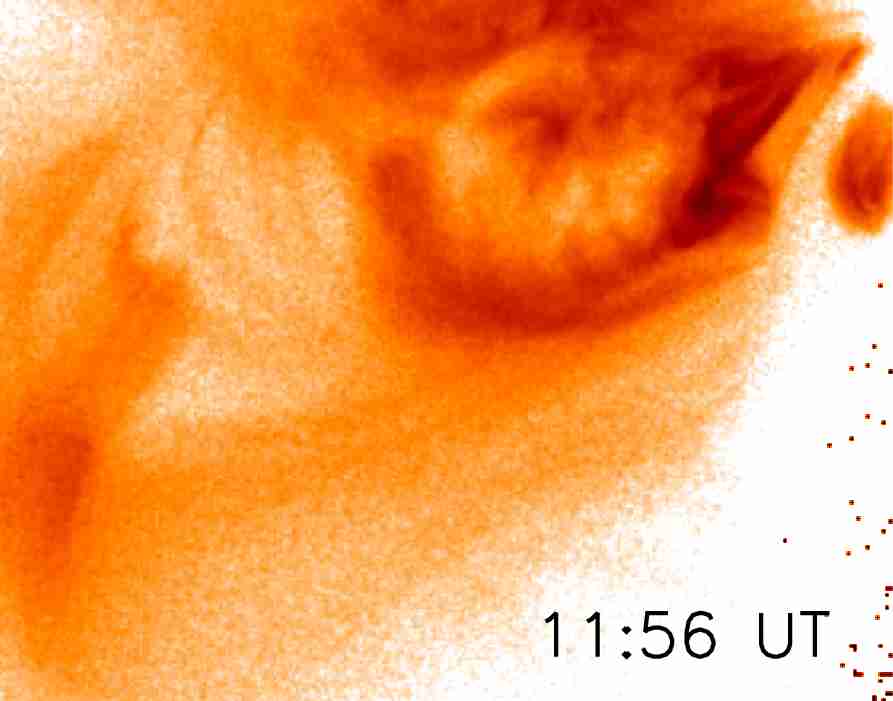}
\caption{Details of the active region showing the evolution in time in the southwest area in the $XRT/Al_{poly}$ filter (and $Be\_thin$ for only 11:26UT).  The first two rows show the region X in the first part of the flare in a reduced FOV of $110\times 110$ arcsec. The last two rows show a larger FOV  ($290\times 220$  arcsec)  also including the loops H and L. The black box marks the FOV of the first two rows. The intensity scale of the images varies in time to highlight the morphology of the loops.}\label{Al_poly_time}
\end{figure*}

Figure \ref{Al_poly_time} shows the time evolution loop systems X, H, and L  between 10:36 UT and 11:56 UT on the 12th November, as observed in the $Al\_{poly}$ XRT filter (to avoid too much saturation the image at 11:26 UT in the $Be\_thin$ filter). The first two rows emphasize the changes in the loop system X, while the last two  look at the whole area (the box in the plot at 11:31 UT marks the FOV of the first two rows).
The figure clearly shows the rapid evolution of the topology of region $\mathrm{X}$ following the flaring events, with a continuous reorganization of the magnetic field.

\begin{figure*}[ht]
   \centering
 \includegraphics[scale=.8]{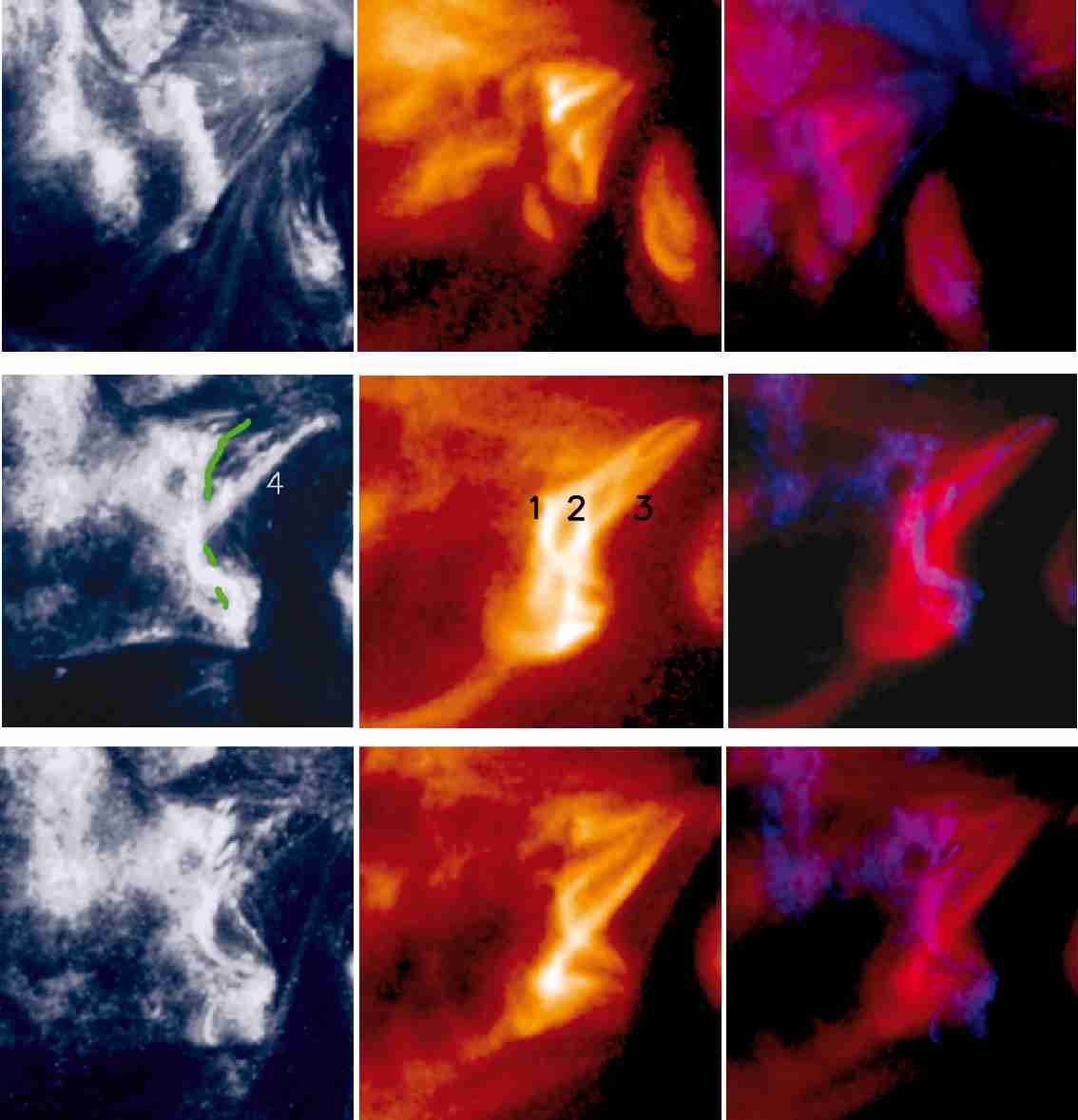}
\caption{TRACE (left) and XRT/Al$\_poly$ (center) intensity maps of the same region as the top two rows in Fig.~\ref{Al_poly_time} (FOV of $110\times 110$ arcsec), and their superposition (right). Top: before the flare (2006 November 10); middle: between the flare peak and the second brightening (11:21 UT); bottom: after the latter (11:56 UT). Some loop structures discussed in the text are labeled with numbers. The green line marks a filament channel.}
             \label{figcompos}%
    \end{figure*}

Prior to the flare, at 10:36 UT, we can distinguish a few loops in the region $\mathrm{X}$, which are  lighted up. They are still  quite aligned. The following image (10:41 UT) taken soon after the beginning of the flare (10:38 UT)  shows at least four loops in a dynamic configuration. One of the loops 
 is no longer aligned with the other and it creates bridges with the neighboring loops.  The loops seem to get closer to a cross-configuration in the central region and near the south footpoint. At this time the image is saturated at the crossing of the loops and at both footpoints.  The former location is the main site of the flare.
In the following frames the flaring region is completely saturated and the flare reaches its maximum.

Later the loops appear to relax to a more simple and aligned topology (10:56 - 11:01 UT), but then other structures get entangled (11:16 UT), all
flare at 11:26 UT and saturate the filters again (Fig. \ref{goes} phase III, and see also second row of Fig. \ref{figcompos}). This kind of  change of magnetic configuration was already reported by \cite{su07}, who noticed a decreasing of the shearing field  after X-Class flares.

By this time, the hardest filters in our data -- which are the only ones not saturated -- reveal that the southern footpoint of the group of loops X is the  brightest.  This location is the same as  one footpoint of the loop system H, which makes it a candidate to be also involved in the flare. At around 11:30 UT - 11:36 UT only the southern footpoint is still saturated, which indicates  that his second brightening was located south of than the main flare.  
 After the second saturation, the small loops in region X appear again in a less complex geometry, even though some bridges betweens loops are still present. This geometry can be seen starting from the images at 11:41 UT in Fig. \ref{Al_poly_time}.

The last two rows of  Fig. \ref{Al_poly_time} show that loops L and H, particularly loop H, are both ignited after the flare.

In the TRACE 171 \AA~ images (Fig. \ref{figcompos}) only one compact loop is clearly visible and bright ten minutes before the second XRT brightening at 11:27 UT (this is possibly a post-flare loop). We also see a filament channel in absorption along the moss, marked in green on Fig. \ref{figcompos}, which is better visible at later times. Its northern part seems to end in the spot while the southern part ends in the loop footpoints. This path corresponds to the inversion line visible in the magnetic map in Fig. \ref{al_p}. 

Figure \ref{figcompos}  shows the intensity maps of the southwest part X of the active region in TRACE 171 and XRT/$Al\_poly$ bands and their superposition respectively. The images of the first row show the  absence of loop-tangling  during the quiet period on the 2006 November 10. In the middle XRT image (11:21 UT) we see three or more loops (numbered from 1 to 3 in the figure) entangled with each other. Their footpoints lie  in the spot center and  south of region X. In the corresponding TRACE image we see a compact bright loop system (loop 4) which, after coalignement and superposition with the XRT image (right panel), appears to be distinct and entangled with the XRT loops. This bundle of loops seems to have a right-handed helical pattern.
Looking at the  evolution  after the flare, we noticed that this TRACE loop did not have an earlier XRT counterpart. On the other hand, these XRT loops never seem to  cool down to TRACE temperatures.
Thirty minutes after the second brightening (bottom row) there are no longer crossing loops, as well as before the flare, even if probably some stressing is still present (more flaring activity was registered later in the day).

\subsection{Temperature}
\label{sec_temp}

\begin{figure}[th]
   \centering
\includegraphics[scale=.6]{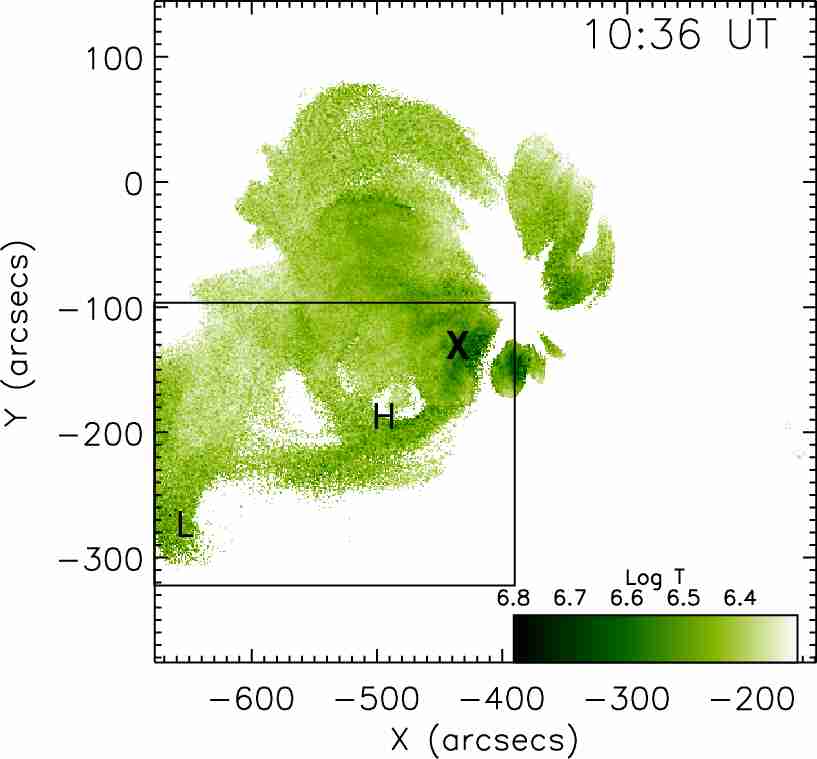}
\caption{Temperature map obtained with the CIFR at 10:36 UT. The color scale is $6.3~<\log~T<6.8$, the darker the hotter. The box indicate the subfield plotted in Fig.\ref{temp}.}\label{temp0}
\end{figure}

 We first study the time evolution of the bulk  temperature. With this quantity we examine the 1-3 MK range, where the active regions mostly emit. 
 Figure \ref{temp0} shows the temperature map of the full active region at 10:36 UT obtained with the CIFR method. The range of the color scale  is $\mathrm{6.3<log~T <6.8}$, with the dark (green in the online version) referring to the hottest regions. The box indicates the subregion plotted in Fig. \ref{temp} bottom, where we made a selection in time to show  the temperature evolution.
The color scale in Fig. \ref{temp0} and Fig. \ref{temp} is the same.
The saturated pixels in the flaring region have been marked in white and were excluded from the analysis because of low photon statistics (see Sect.~{\ref{sec_top})}. We estimated an average error of $\sim 6\%$ as the standard deviation of the temperature.
The noisy aspect of some of the maps is due to their extremely short exposure time (see Table \ref{table1}). 

 The  map at 10:36 UT shows a low and almost uniform temperature before the flare,  except for hotter loops in region X, which later will be the location of a series of flares. Below we concentrate our attention on the marked subregion that is the only one which is affected  by the flare.

The thermal evolution of the area is clear. At the beginning of the sequence the heated region is  very localized, while later on it spreads to the whole area. 
The time sequence of the temperature maps {in Fig. \ref{temp} shows that after the main flare the hot region spreads from the flare's core  to loop H first and then to loop L. 

At 10:41 UT the filters start to saturate and 
at 10:46 UT  a large region is saturated by the flare.
The unsaturated boundary of the core zone has  an average temperature of $8  \times 10^6~ \mathrm{K}$. 
 This temperature is quite low  for flare standards ($T \geq 10$ MK), but we should consider that we are outside of the core of the flaring region, we are outside of the flare maximum, and the filter ratio is soft and weights the cooler components more  ($T \sim 3$ MK). It is however remarkable that this soft filter ratio is able to measure a significantly higher temperature, expected for a flare. This provides a consistency check for filter diagnostics and calibration.
By 10:46 UT, the long loop system L has not been heated perceptibly yet, while 
the other loop system $\mathrm{H}$  has visibly heated up to $\approx 4 \times 10^6~ \mathrm{K}$.

The first unsaturated data  after the flare are available at 11:01 UT. These data were taken with very short exposure times, and we can recover information only in the core of the flaring region and loop H. The image shows that some of the short loops in region X have already cooled down.

\begin{figure*}[th]
   \centering
 \includegraphics[scale=.35]{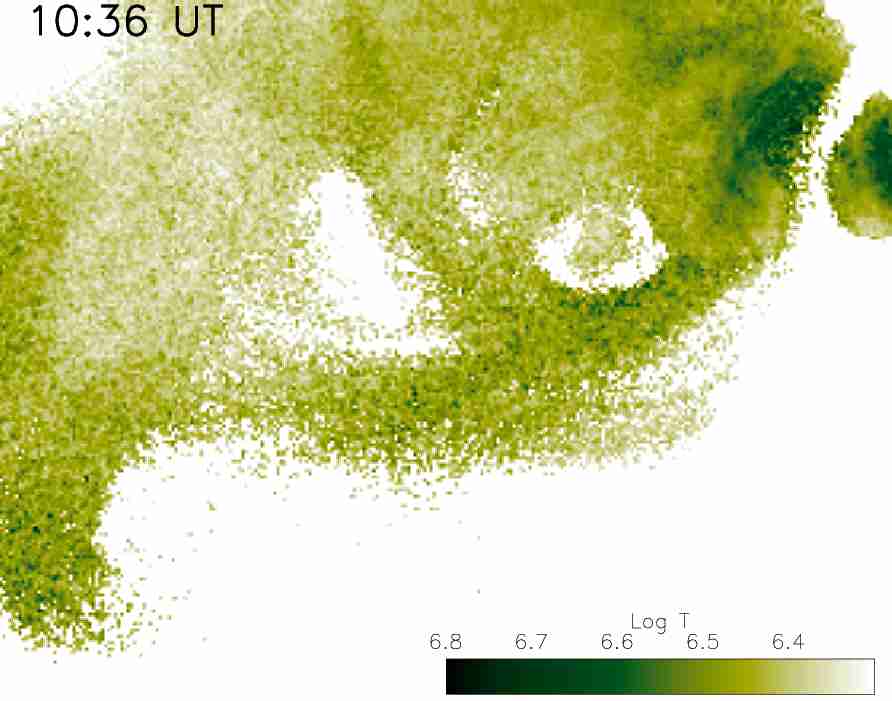}
 \includegraphics[scale=.35]{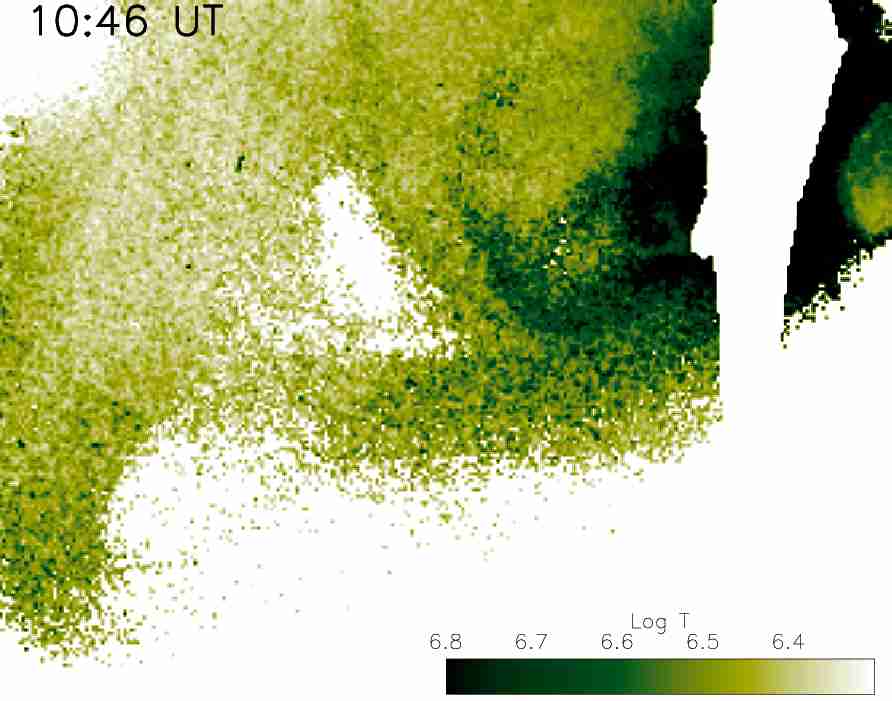}
 \includegraphics[scale=.35]{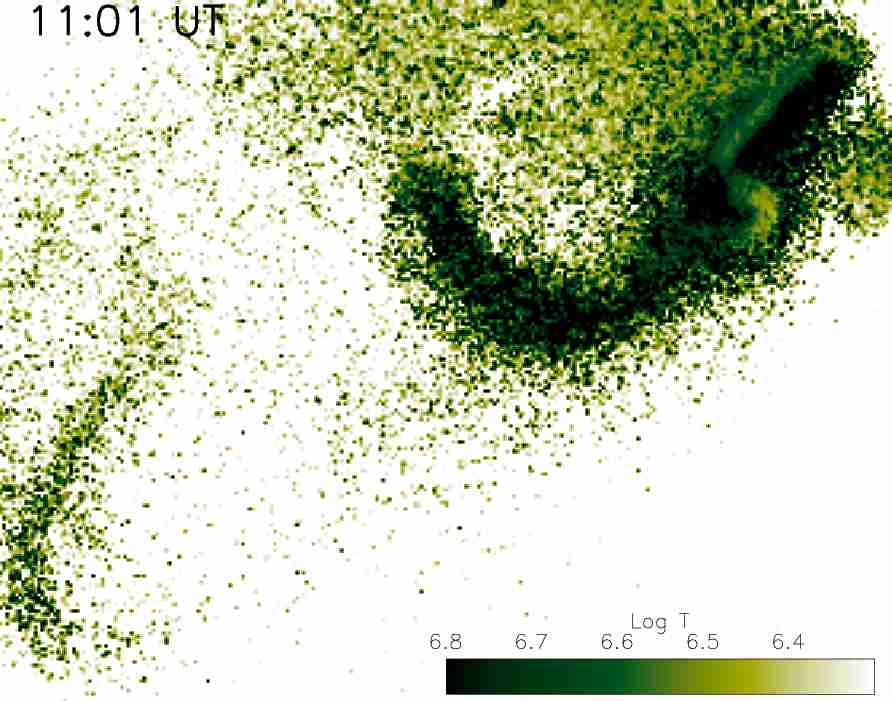}\\
 \includegraphics[scale=.35]{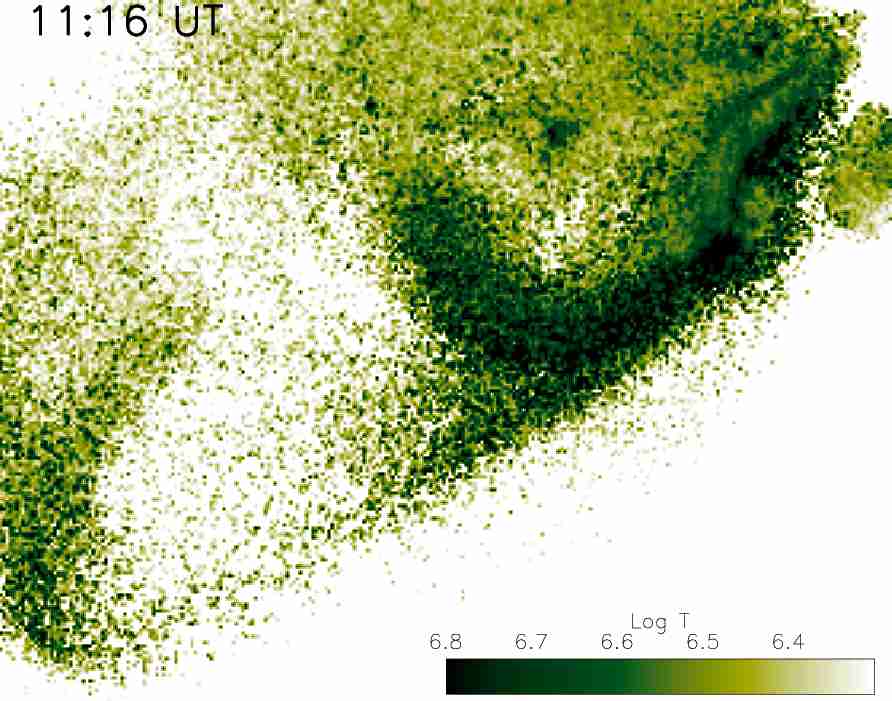}
 \includegraphics[scale=.35]{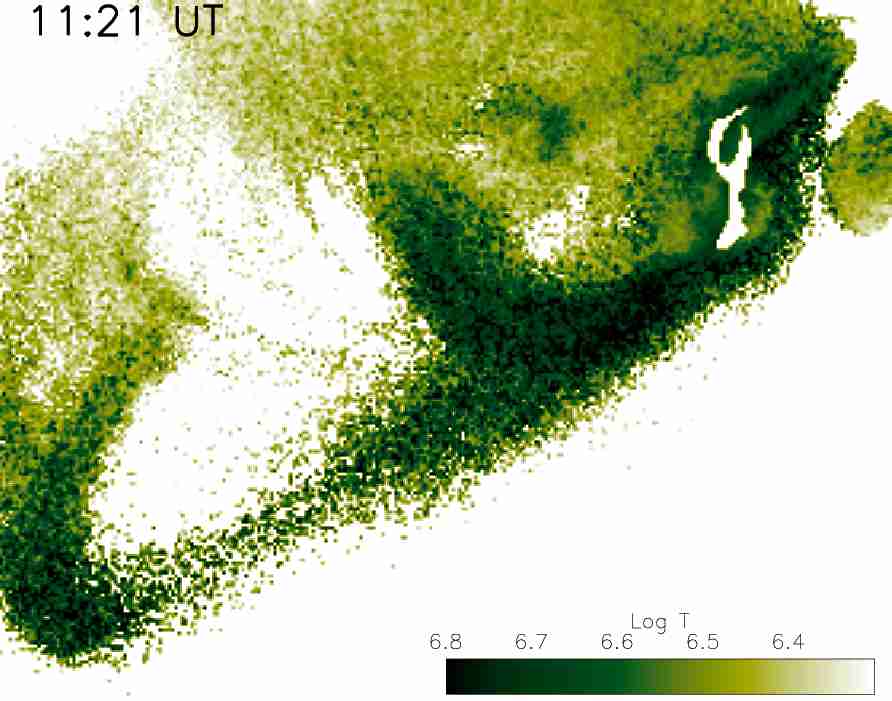}
 \includegraphics[scale=.35]{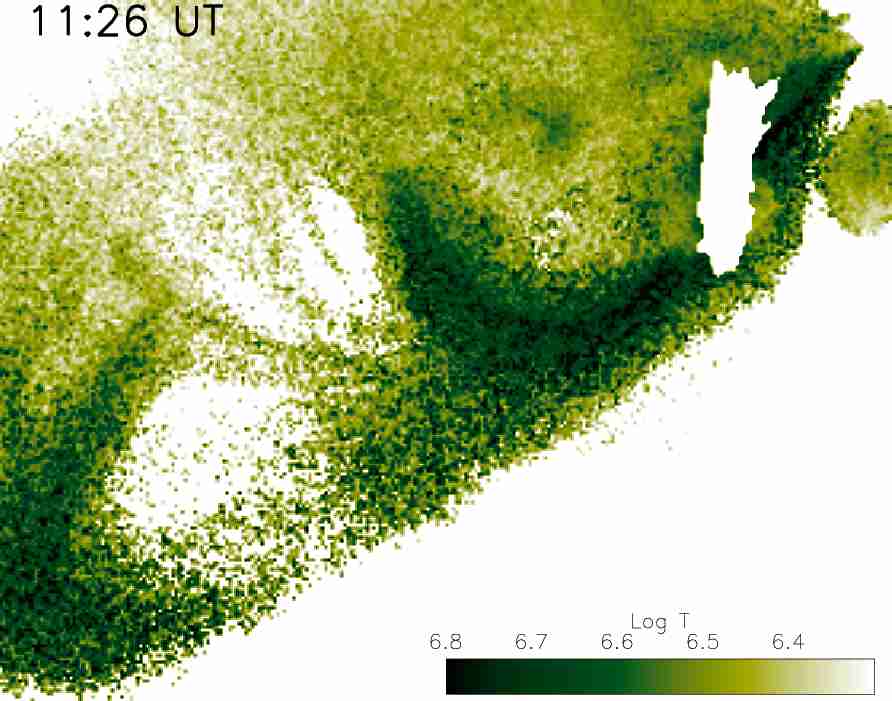}\\
\includegraphics[scale=.35]{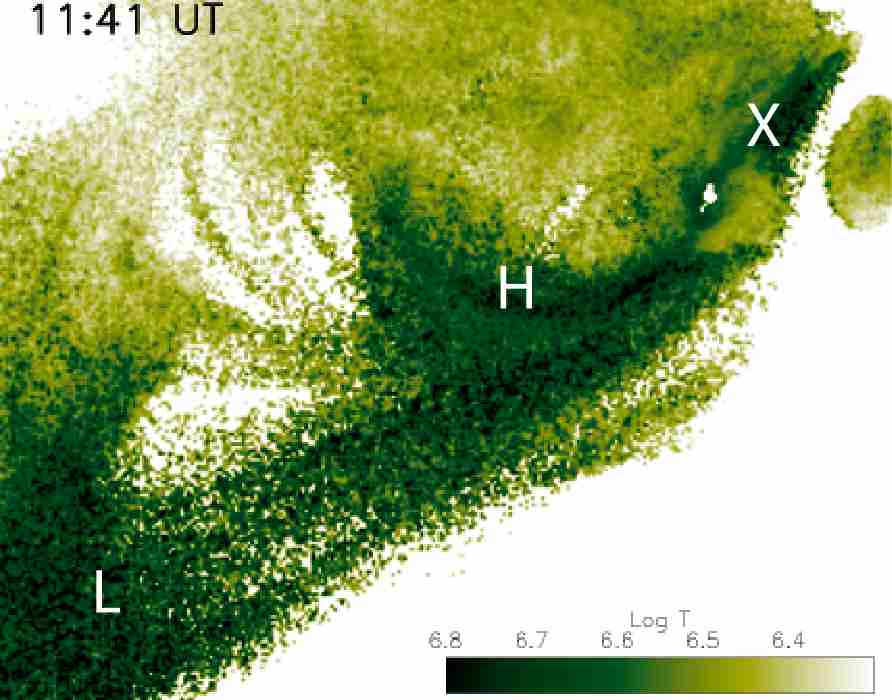}
 \includegraphics[scale=.35]{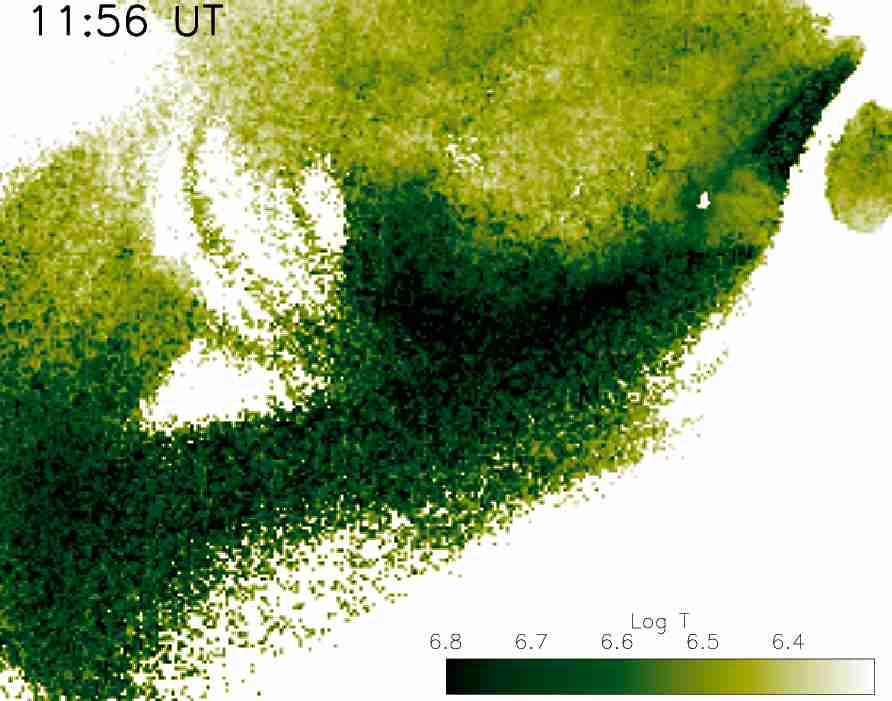}
 \includegraphics[scale=.35]{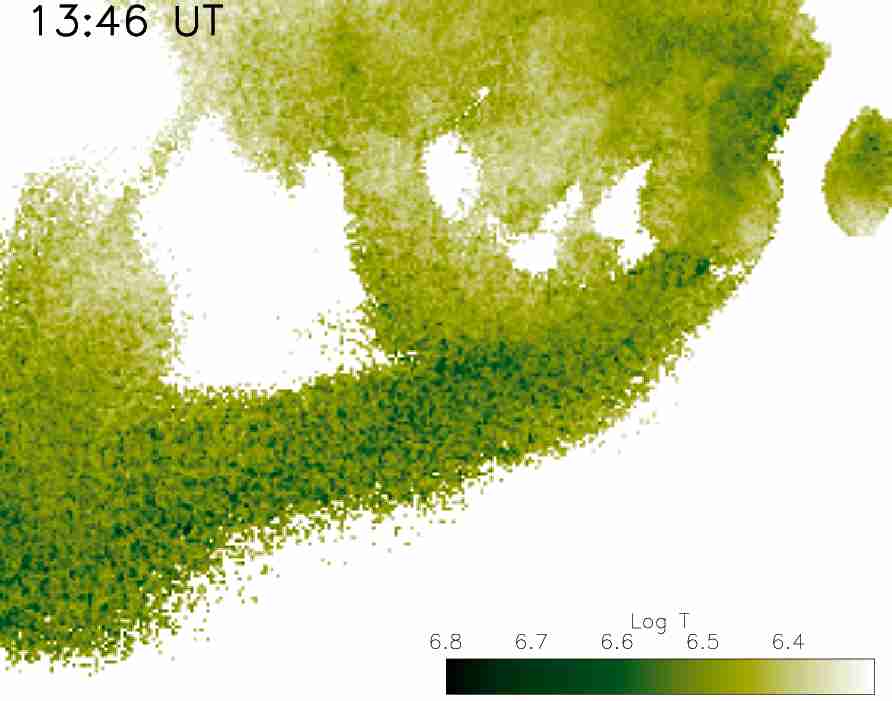}
 \caption{Selection of temperature maps obtained with the CIFR of the subregion marked in Fig. \ref{temp0} (FOV of $290\times220$ arcsec). The color scale is $6.3~<\log~T<6.8$, the darker the hotter.
}\label{temp}
\end{figure*}

At 11:21 UT only a small part of the X area is again saturated, and we can better resolve 3-4 loops (see also the  detailed loop system in Fig.  \ref{figcompos}). 
What appears as a twisting of similar loops in Fig. \ref{figcompos} (middle row) results in different temperature structures in Figure \ref{temp}.
After 11:01 UT one of the loops is clearly cooler than the others, even if it is not clear which one.
At 11:21 UT this can be identified as the western one ($\#$ 3) of the three visible loops in the unsaturated intensity map. 
This is  surrounded by hotter plasma of the other loops of region X.
However, it is interesting to notice that loop $\#$3 is not cool enough to be seen in the TRACE filter (see also Fig. \ref{figcompos}).

Figure \ref{temp} clearly shows that 
the high temperature plasma is expanding from a localized area to 
the longer southern loops of the active region, while the shorter loops in X area are cooling down.
At the end of the time sequence (13:46 UT) the plotted area looks homogeneous in color and darker than at the beginning (10:36 UT). This seems to indicate that the area has a higher temperature than at the beginning, which is more uniform than during the flare. The loops are not thermally distinguishable from the rest and their average temperature is $\approx 3 ~\mathrm{MK}$.

We  measured variations of temperature in the rest of active region that may arise from the flaring activity. To do this we  selected three box regions A and B and C of about $40 \times 40$ pixels (Fig. \ref{al_p}). The temperature measurements from the beginning to the end of our time sequence show that these regions have constant temperatures of about $2.6 ~\mathrm{MK}$. This confirms that the flare acts only locally.

Figure \ref{fig_t_em_time} (left)  quantifies the temperature variations in the different structures discussed here. For these measurements we specifically selected the pixels in  boxes near the top of loops H, L and in the flaring area X near the loop footpoints (see boxes in Fig. \ref{al_p}). 
The latter was identified as the best place to measure the temperature as close as possible to the flare and which has the minimum number of saturated exposures.
The fluxes were integrated in the boxes to derive the average temperature and the total emission measure (discussed in the next section).

 In region X the flare determines a temperature peak at about 10:55 UT. The temperatures derived from the CIFR method  put this peak at around 9 MK, which is to be interpreted as a weighted average of a non-uniform emission measure distribution along the line of sight and as an effect of excluding the hottest but saturated core. The temperature of loop H also clearly increases and shows a peak a few minutes later than region X. Loop L shows a more gradual temperature increase, which is still going on at the end of the time sequence. Curiously, after 11:16UT all selected regions settle to a similar temperature level ($\sim 5$ MK).

In a flare analysis better diagnostics are expected from the ratio of the hardest filters. Unfortunately, our attempt to derive temperature maps from the hardest  ratio available (F4/F5) was not  successful, because we could not find a trade-off between binning to improve the $S/N$ ratio and keeping spatial information. Binning over boxes of 4x4 pixels was not enough to significantly improve the $S/N$ ratio  to obtain a relevant map. Binning over larger boxes mostly cancels out the spatial information in the saturated flaring region.
However, F4/F5 thermal maps are useful for the rest of the active region, as  will be discussed in the next section. 
We remark that by means of the CIFR method we were  able to obtain both temporal and fine spatial information about the temperature structure of the flare region.

\begin{figure*}[ht]
   \centering
\includegraphics[scale=.5]{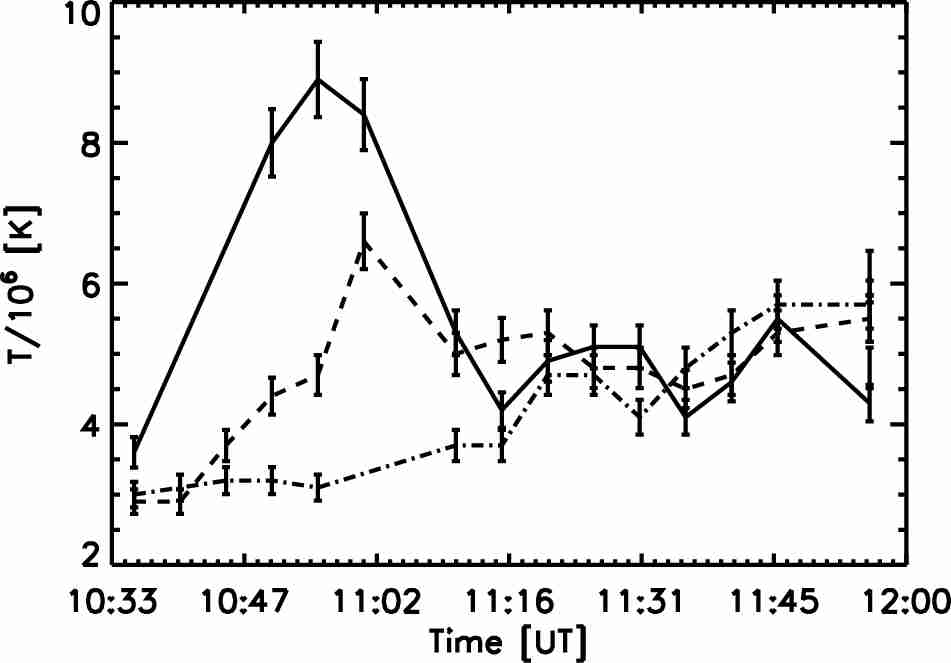}
\includegraphics[scale=.5]{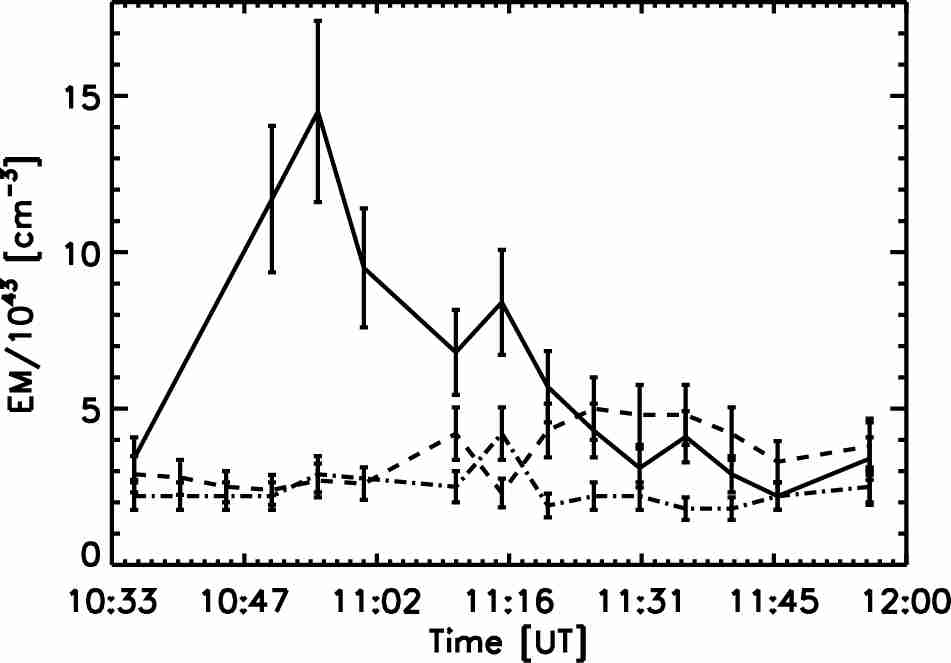}\\
\caption {Evolution of the average temperature (from CIFR) and emission measure for the top of loops L (dashed-dotted) and H (dashed), and the box in region X (solid). 
Note that some of the measurements are missing (see text). }\label{fig_t_em_time}
\end{figure*}

\subsection{Emission measure}
\label{sec_em}

Once the temperatures are known from the filter ratio, it is possible to derive the emission measure in a pixel. This calculation is still an estimate because  it is done under the assumption of
isothermal plasma along the line of sight. 

Figure \ref{em_map} shows the EM maps of the same region as Fig. \ref{temp} before and after the flare, derived assuming the temperatures from the CIFR. For these maps we estimated an error of about $20\%$ in the pixel.
In the maps we can identify part of the denser core of the active region which includes the flaring loops. The larger structures are less dense even if we see an increase  of the EM after the flare in loops H and L.

\begin{figure*}[ht]
   \centering
 \includegraphics[scale=.4]{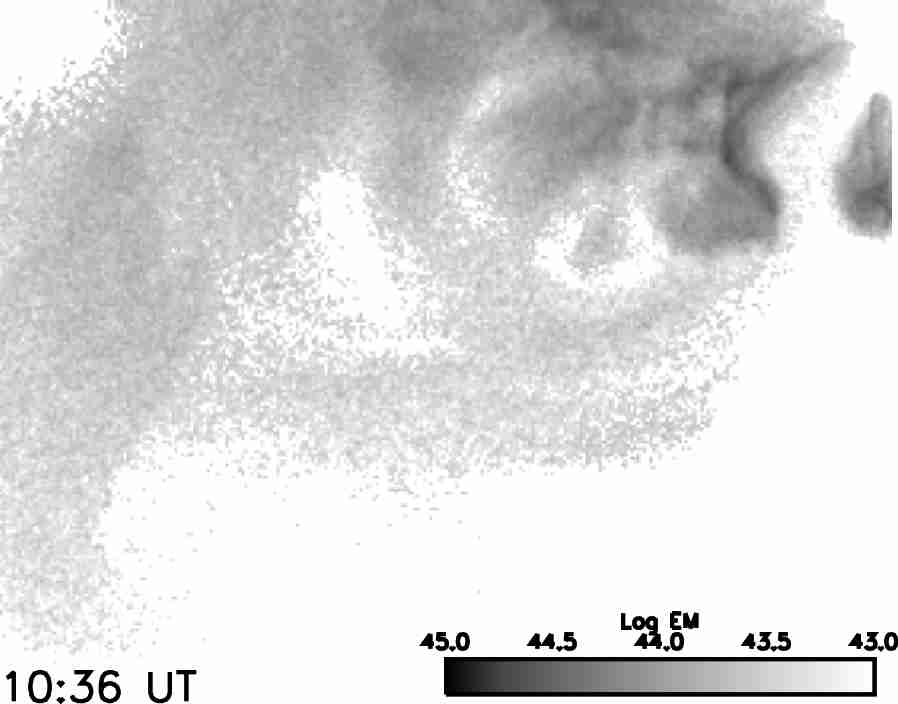}
 \includegraphics[scale=.4]{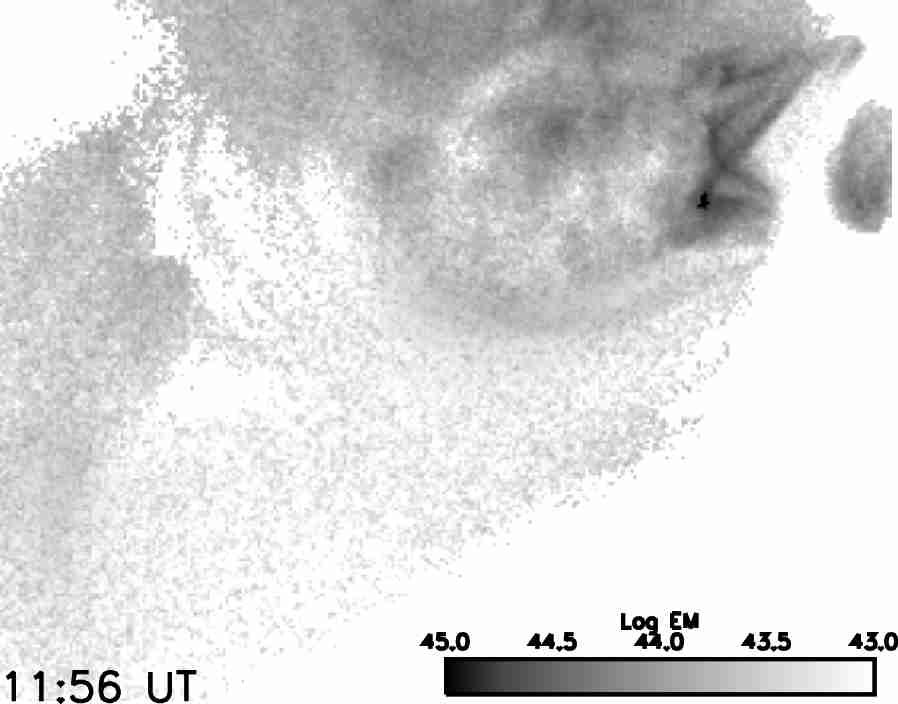}
\caption{Emission measure maps before and after the flare derived using the temperatures from the CIFR  (the FOV if the same as that of Fig. \ref{temp}: $290\times 220$ arcsec).}\label{em_map}
\end{figure*}

The information contained in these maps together with those from the temperature were used to estimate how the EM (we are still under the isothermal assumption inside a pixel) is distributed in temperature in the whole selected area. This is possible through the emission measure $vs$ temperature diagram (EM(T), e.g. Orlando et al. (2000)), built by summing up the EM values in each temperature bin.
In addition, by plotting on the same diagram the EM values found from filter ratios sensitive to different temperatures, it is also possible to recover some multithermal information. 

Figure \ref{emt_plot} shows the EM(T) plots for the values derived with the CIFR and F4/F5 for the whole selected area of the active region at different times of the flare. The values were normalized to the total EM in each map  to better compare the values derived from the different filter ratios.
In general we found the results from the two ratios to be consistent in the temperature range where they overlap, confirming the validity of the diagnostics and the overall consistency of the analysis.

From these maps we see that prior the flare the plasma was distributed almost homogeneously around $\log~T= 6.35$, without hot plasma from the hardest filter ratio. At the time of the flare's peak (10:46 UT), besides the saturated  core pixels, the EM distribution starts increasing on the high temperature wing, with a small component of plasma at 10 MK. 
The first unsaturated image in the sequence is at 11:01 UT. Here the EM is more broadly distributed in the high temperature tail, although the quantitative agreement between the two filter ratio is worse around $\log T \sim 6.8$. This is probably due to the photon noise for these low exposure times (the error in the F4/F5 ratio in this exposure may reach about $50\%$).

One hour after the main flare, the EM
distribution still shows a high temperature tail, even though with less amplitude. Hot but faint plasma is still detected through the F4/F5 ratio. The F4/F5 ratio appears invariably to detect additional hotter components with respect to the other filter ratio. Although its presence is expected in a flare, we cannot conclude whether this hot component is real even in part, or produced by the photon noise (Reale et al. 2009).
In any case, the results shown in the figure  point out the global and quantitative consistency of the information obtained from different filter ratios.

\begin{figure*}[ht]
\includegraphics[scale=.31]{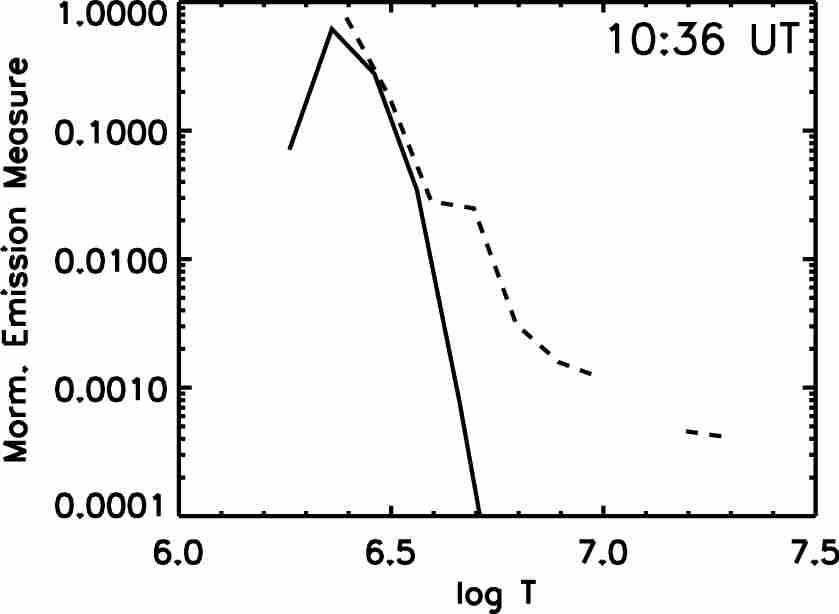}
\includegraphics[scale=.31]{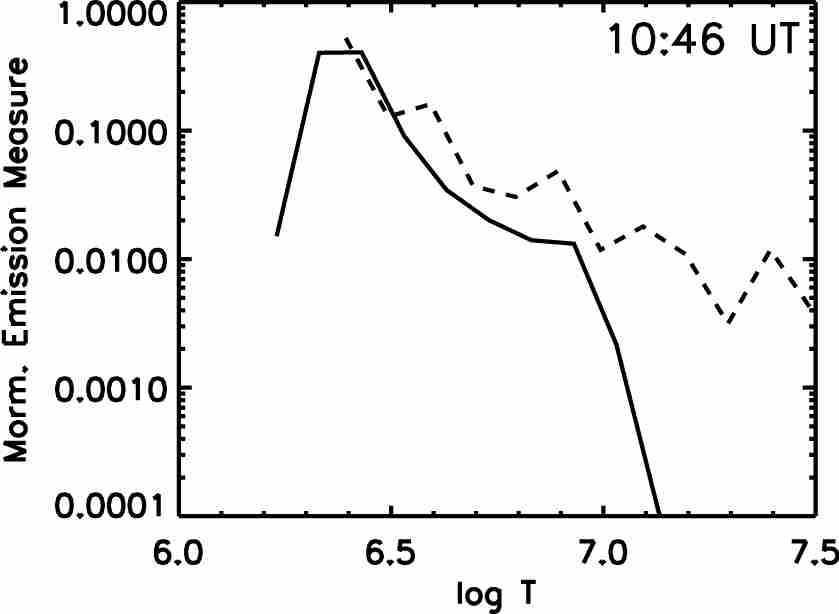}
\includegraphics[scale=.31]{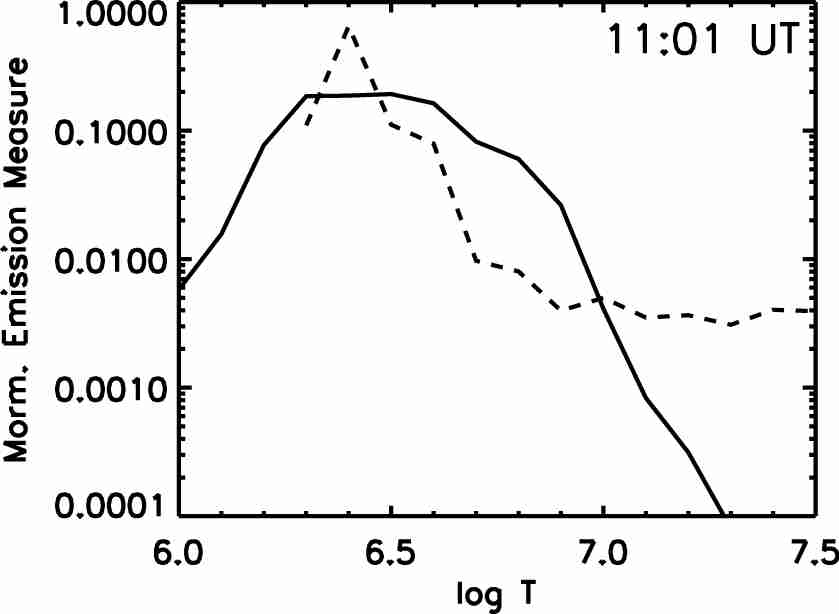}
\includegraphics[scale=.31]{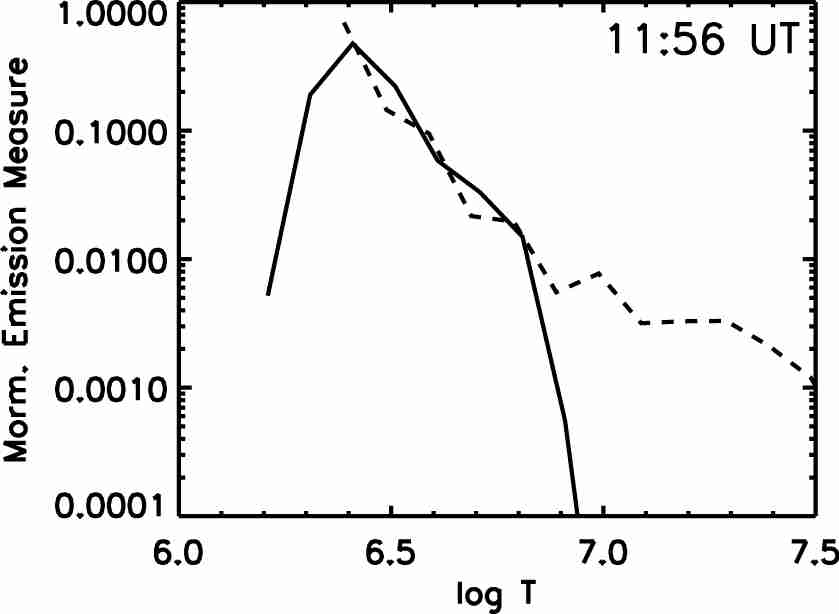}
\caption{EM(T) plots for different phases of the flare. The solid line refers to the EM derived from the CIFR, while the dashed line is from the F4/F5.}\label{emt_plot}
\end{figure*}

The evolution of the EM(T) distributions seen in Fig.~\ref{emt_plot} made us to look back in more detail at the evolution of the EM in specific areas of the loops, shown on the right of Fig. \ref{fig_t_em_time}. There the EM is the average per pixel, in order to make boxes with different areas comparable. The variation of the EM in the X region follows the flare evolution. There is apparently no time delay between the peak of the EM and the temperature, while it is there for loop H. The variation of EM is not significant in loop L. If we assume that the volume in the boxes does not change with time, the variation of the EM is a signature of the variation of the plasma density. In this case Fig. \ref{fig_t_em_time} indicates that between the beginning and the end of the flare there has been an increase of pressure.
While the chromospheric evaporation is present in the X region and probably to some extent in loop H, this is not the case for loop L.

\section{Discussion}
\label{sec_disc}


 We analyzed the evolution of a flare in a multi-filter observation by Hinode/XRT.
The results highlight rapid morphology changes on XRT data (at 5 minutes cadence) of the shorter loop system  (region $\mathrm{X}$) directly involved in the flare. The changes consisted in systematic cycles of first tangled, then  relaxed configurations, with the maximum of the tangling just before a flare.
The combination with TRACE data shows small changes at 1MK with only a bright isolated loop bundle placed in the flaring region. The fine-scale analysis possible with these instruments shows the TRACE bundle to be  tangled with the XRT loops just before a brightening of the flare.  

Our results are consistent with the picture where the energy released during the flare seems to come from the magnetic tangling of loops in corona. The conversion in plasma heating is confirmed by our thermal analysis of the area.  
We derived temperature maps 
using  the five available filters of XRT and applying the new CIFR method (Reale et al. 2007). 
The data show the propagation with time of the heated structures from the flare core to intermediate ($\mathrm{H}$) and then larger loops ($\mathrm{L}$). Figure \ref{temp} seems to show that both loops are heated asymmetrically starting from their western footpoint (see map at 10:46 UT for loop $H$ and 11:16 UT for $L$). 
Given this and the time and space proximity of the temperature increases, we conclude that there is a clear correlation between the flare and the ignition of the other two loops. The main flare is accompanied by secondary heating, which involves nearby and larger loop structures. The larger size of the latter structures and the smaller amount of energy released in them make their brightening far less conspicuous than the main flare. This may be an evolution similar to that which leads -- on far larger energy scales -- to the ignition of arcade flares  \cite[e.g][]{aschwanden01b}.

Using the EM information  we estimated the total energy of the loops $H$ and $L$ required  to reach the measured temperatures and its weight on the total flare energy budget.
Due to  saturated pixels it was not possible to achieve an accurate estimate of the total energy of the flare at the moment of its peak. Instead we deduced the measurable part of it from the radiative losses obtained using the first unsaturated images at 11:01 UT. We determined the total EM in the flaring area (which includes the group of bright loops visible in Fig. \ref{Al_poly_time}) to be $7\times 10^{47}~ \mathrm{cm^{-3}}$ and the average temperature to be $5.4\times 10^6 ~\mathrm{K}$. Using the radiative loss curve from the CHIANTI $v.6$ database \citep{dere09} and the e-folding time (600 sec) of the GOES curve (Fig. \ref{goes}), we obtained a total energy of $3.6\times 10^{28}~ \mathrm{erg}$. 

For the other two loops the EM before the flare is not negligible with respect to that during the flare, which is different from the proper flaring loop $X$. Accordingly the radiative losses from the other two loops were obtained by measuring the increase of total EM (integrated in all the loop) from 10:36 UT to their maximum, that is 11:16 UT  for loop $H$ ($\Delta EM =4.8 \times 10^{46}~ \mathrm{cm^{-3}}$) and 11:56 UT for loop $L$ ($\Delta EM =10^{46} ~\mathrm{cm^{-3}}$). With these values we obtained a budget of $8 \times 10^{27} ~\mathrm{erg}$ for loop $H$ and  $2.9 \times 10^{27}~ \mathrm{erg}$ for loop $L$. 
These results show that while the energy required to heat loop $L$ is a small fraction of the flare's energy, loop $H$ requires about $1/4$ of it. However, in both cases this energy is not enough to bring each of these loops to flare, as confirmed by Fig. \ref{fig_t_em_time}.

The average flare temperature and total EM derived from XRT data at 11:01 UT can be compared to the values derived from the GOES data. From the GOES filter ratio we obtained $T \approx 8.5 ~ \mathrm{MK}$ and EM $\approx 3\times 10^{47}~~ \mathrm{cm^{-3}}$, which are close to those found with XRT. Considering that GOES is sensitive to higher temperature than XRT, these values are consistent with the results of \cite{mctiernan09}, where the EM decreases as
the temperature increases. 

From Fig. \ref{figcompos} we see that the XRT and TRACE loops coexist sharing the same loop footpoints. While the XRT loops are continuously seen at least for all the duration of our data, the TRACE loop almost disappears around 11:35 UT and is not bright again for several hours after the flare. We  checked the relative position of the two bundles of loops up to this time and noticed that before disappearing the bright TRACE loops become part of the western XRT bundle. As shown by our results, all the analyzed loops at 12 UT are hotter than before the flare (Fig. \ref{fig_t_em_time}). It is possible that the TRACE loops were also heated to the point where they almost disappear from the TRACE band. 

In conclusion we showed that  it is possible to make diagnostics of temperature, EM and dynamic analysis over  fine spatial scales and relatively short time scales using XRT data.

\begin{acknowledgements}
      SP acknowledges the support from the Belgian Federal Science Policy Office through the ESA-PRODEX programme. FR acknowledges support from Italian Ministero dell'Universit\`a e Ricerca and Agenzia Spaziale Italiana (ASI), contracts I/015/07/0 and I/023/09/0. KKR is supported by NASA contract NNM07AB07C to SAO. 
The authors acknowledge the anonymous referee for the useful suggestions on the manuscript.
Hinode is a Japanese mission developed and launched by ISAS/JAXA, with NAOJ as domestic partner and NASA and STFC (UK) as international partners. It is operated by these agencies in co-operation with ESA and NSC (Norway).

\end{acknowledgements}

\bibliographystyle{aa} 
\bibliography{bib}
\end{document}